# Hidden structure in amorphous solids

F. Inam[1], James P. Lewis[2] and D. A. Drabold[1,3]

[1] Department of Physics and Astronomy, Ohio University, Athens, Ohio 45701, USA

[2] Department of Physics, West Virginia University, Morgantown, West Virginia, 26506, USA

[3] Trinity College, Cambridge, CB2 1TQ, UK

Recent theoretical studies of amorphous silicon [Y. Pan *et al*. Phys. Rev. Lett. **100** 206403 (2008)] have revealed subtle but significant structural correlations in network topology: the tendency for short (long) bonds to be spatially correlated with other short (long) bonds. These structures were linked to the electronic band tails in the optical gap. In this paper, we further examine these issues for amorphous silicon, and demonstrate that analogous correlations exist in amorphous $SiO_2$, and in the organic molecule, $\beta$-carotene. We conclude with a discussion of the origin of the effects and its possible generality.

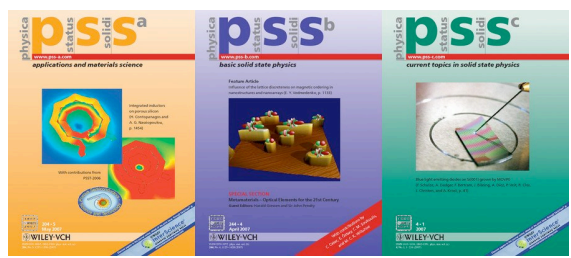

Selected covers of pss (a), (b), and (c) as an example for the optional abstract figure (with or without caption). If there is no figure here, the abstract text should be divided into both columns.

**1 Introduction** The continuous random network (CRN) has become the accepted model for homogeneously disordered materials, and for good reason: such atomistic models reproduce the experimentally observed properties of the materials, including some of the finer details, such as the Urbach tailing in the electronic density of states[1][2][3][4].

Recently [5], we have closely examined some particularly realistic CRN models of amorphous silicon (a-Si), and reported that there are subtle correlations that have not been previously noted: short bonds tend to be connected to other short bonds and an analogous statement for long bonds. In a-Si, the longer bonds form filamentary (1-D) structures, and the short bonds tend to cluster in a higher dimensional way, albeit with some 1-D character. We have shown that the Urbach tails in the electronic density of states arise from these structures [4](the valence tail associated with the short bonds, conduction tail with long bonds).

The purpose of this paper is to explore the case of a-Si a bit farther, and to look for related phenomena in glasses (silica) and beyond. Obvious questions include: (1) Are "filaments" universal? (2) Do these topological features offer a general explanation for the origin of the Urbach tails? (3) Why do these structures form, and can they be manipulated to tailor the band tails in a useful way? We do not offer complete answers to these questions, but provide new information relevant to all three.

**2 Methods** Where a-Si is concerned, we focus our attention on a 512-atom model due to Djordjevic, Thorpe and Wooten (DTW)[6], made using the Wooten-Weaire-Winer (WWW) approach[1]. The DTW model is thoroughly relaxed within the WWW scheme, and to our knowledge, is not in significant contradiction with any experiment (structural, electronic or vibrational). The model has perfect four-fold coordination. We have relaxed the model with an accurate *ab intio* code and find tiny rearrangements, with no bond breaking or formation. Where subtle or "hidden" structural correlations are concerned, we have discussed this (and eight other models made in various ways) in detail elsewhere [5].

We look for analogous structural and electronic correlations in a-$SiO_2$. For a-$SiO_2$, we start with a 648-atom

Decorate and Relax model in excellent agreement with experiment[7].

All electronic structure and force calculations are implemented with SIESTA, a standard first principles, local basis code. We updated the model of amorphous silica by relaxing it with a highly optimized double-zeta-polarized basis in SIESTA, kindly provided by E. Artacho. Changes were small compared to the published model.

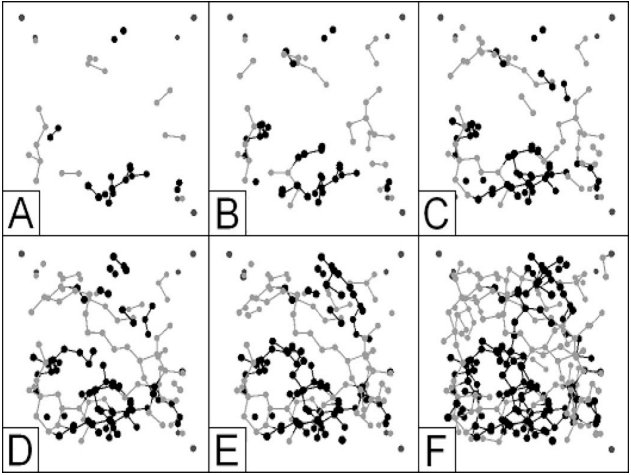

**Figure 1** (A) 1%, (B) 2%, (C) 3% (D) 4% (E)5% and (F) 8% longest (light) and shortest (dark) bonds in 512-atom DTW model[6] of a-Si (from [5].)

**3 a-Si: short recapitulation** Previous work[4][5] on a-Si demonstrated that: (1) long and short bonds are spatially correlated, short favouring proximity with short, and long with long. Furthermore, the long bond structures are filamentary, whereas the short bonds tend to cluster in 3-D (see Figure 1). There is little or no cross-correlation (long bond to short bond); (2) The electronic valence tails arise from short bonds and the conduction tails from long bonds; (3) Energies close to the Fermi level ($E_f$) are localized on the longest (for energies slightly above $E_f$) and shortest bonds (for energies slightly below $E_f$); (4) If the network is changed to eliminate the short-short and long-long correlations (but no bonds are broken or formed) the model does *not* yield a exponential density of states; (5) Simulations of relaxed point defects (Si divacancy) in diamond show a filamentary structure upon relaxation, and hint at the formation of an exponential tail, in apparent agreement with ion-bombardment experiments, that reveal the appearance of an Urbach edge before amorphization [4].

**4. a-Si: Defect nuclei** To quantitatively characterize the long or short bonds and their associated strain or relaxation field, we look at how bond lengths vary as a function of distance from the defect centre or "nucleus" – by this we refer to a central bond smaller (or larger) than any nearby. In Figure 2, we show the characteristic length of a particular short-bond cluster. $\Lambda$ is the longest distance to sites with a normal bond length. We repeat this analysis for many defect nuclei.

In Figure 3, we plot $\Lambda(\delta r)$, where $\delta r$ is the deviation from the mean bond length in a-Si (about 2.35Å). The plots are noisy (and not to be taken seriously in any case for $\delta r/f \ll 1$, where $f \sim 0.09$Å is the FWHM of the a-Si bond length distribution function), but show that for short bonds, $\Lambda$ becomes large; whereas for bonds longer than the mean, there is no obvious functional dependence: long bonds do not imply a long-range relaxation field.

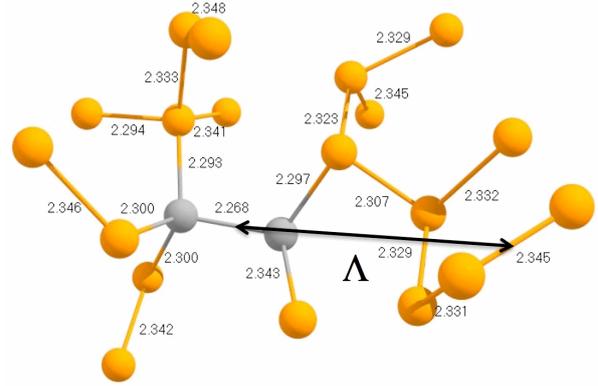

**Figure 2 (Color online) Example of a short-bond cluster in a-Si. $\Lambda$: linear extent of densified region induced by defect nucleus (grey bond).**

We further differentiate between the long- and short-bond clusters by computing the number of atoms $N_{cl}$ participating in the deformation caused by a particular nucleus (Fig. 4). Here, the difference between long and short bond cases is stark, reflecting the strong correlation between the relaxation around the short bond defects and the bond length deviation at the nucleus. This implies that in a-Si, the short bond is a seed for growing a local volume of higher density, and the shorter the bond the larger the effect. Roughly speaking, this suggests that if one is able to create a short bond in any fashion, one can expect the network to locally densify around that nucleus.

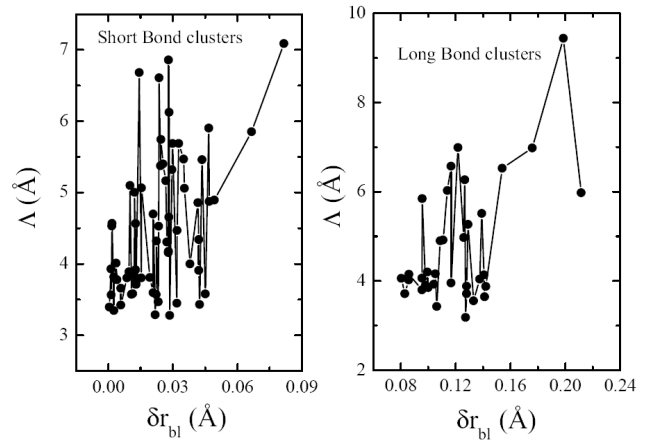

**Figure 3** Plot of linear extent of relaxed defect $\Lambda$ as a function of $\delta$, difference in mean bond length. $\delta<0$ implies short bonds, $\delta>0$ long.

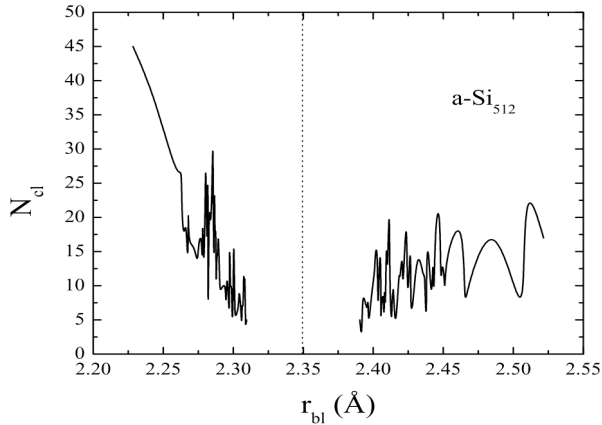

**Figure 4** Number of atoms in short-bond and long-bond clusters ($N_{cl}$) are plotted agianst the central bond length of the cluster for a-Si. Mean bond length is represented by the dotted line.

**5. The case of a-SiO$_2$** The 648-atom a-SiO$_2$ model is chemically ordered, all Si are four-fold, and all O are two-fold. The neutron structure-factor obtained from the model is in excellent agreement with experiments (as are bond angle distributions and other observables). In this section, we explore the network connectivity of amorphous silica using the methods of the previous sections applied to a-Si. In contrast to a-Si, no topological correlations were found for the Si-O nearest-neighbor bond *lengths*. We then considered second-neighbour distances $R_{O-O}$ (with mean length 2.66 Å) and $R_{Si-Si}$ (with mean length 3.05 Å). Since the variation in Si-O bond length is tiny (FWHM ~ 0.01Å), the fluctuation in $R_{O-O}$ and $R_{Si-Si}$ is really due to broadening associated with O-Si-O and Si-O-Si *angles*. The fluctuation in $R_{Si-Si}$ (FWHM ~ 0.2Å) is larger than that of $R_{O-O}$ (FWHM ~ 0.1Å). Fig. 5 indicates a linear dependence between the two lengths. The O sites with shorter $R_{O-O}$ have shorter $R_{Si-Si}$, so that smaller $R_{O-O}$ implies higher densities.

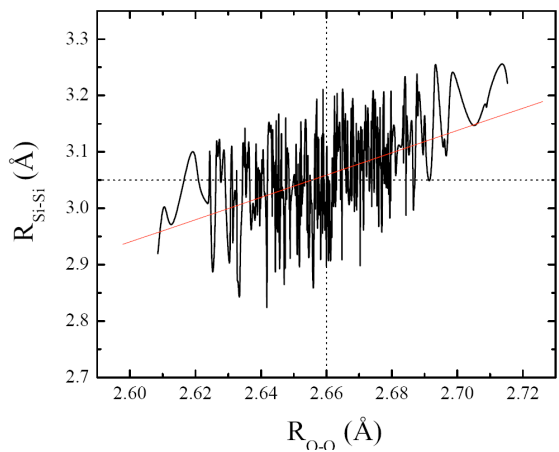

**Figure 5** Correlation between the two lengths $R_{Si-Si}$ and $R_{O-O}$. Diagonal line is a linear fit to the data. Dotted lines indicate the mean values of the two lengths.

In analogy with our work on a-Si (Fig. 1), we show examples of the connectivity of the subnetwork formed by long and short second neighbors in Fig. 6. Fig. 7 shows the correlation between $N_{cl}$ and the short (longer) length clusters of the two kinds. A crudely linear correlation is evident for short $R_{O-O}$, a trend analogous to the a-Si network. Cluster size $N_{cl}$ also exhibits linear behaviour for large $R_{Si-Si}$. The connectivity of short $R_{O-O}$ clusters tends to be of a compact 3-D nature, while large $R_{Si-Si}$ clusters are more extended and filamentary (1-D). The average spatial extension of the small $R_{O-O}$ clusters is of order 5.0Å reflecting the denser nature of the surrounding network induced by the defect seed or nucleus.

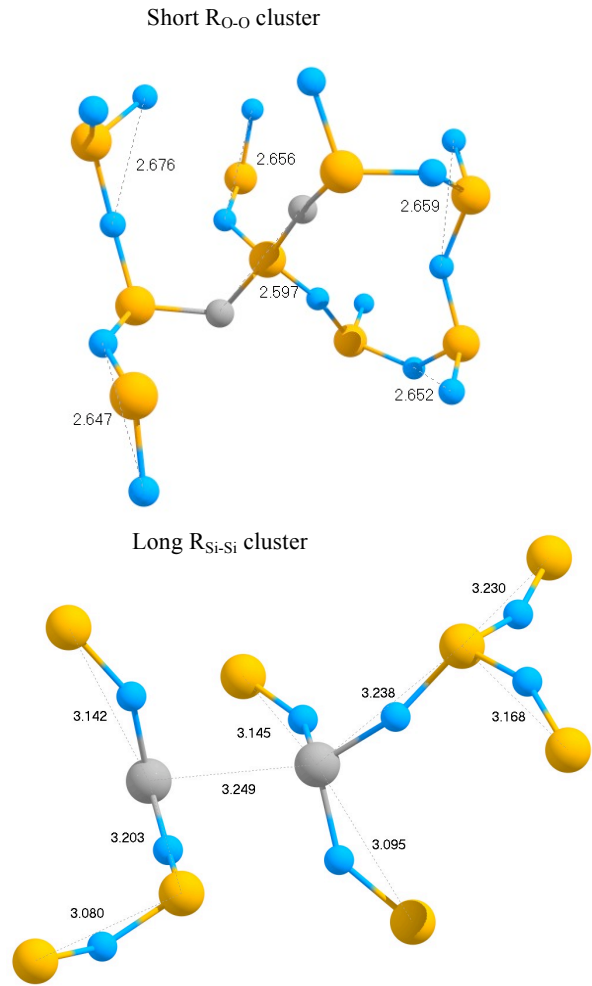

**Figure 6** (Color online) Examples of short $R_{O-O}$ and long $R_{Si-Si}$ clusters in a-SiO$_2$ network. Grey sites represent the defect nuclei.

The link between topology of the network and the electronic spectrum is analyzed by calculating the symmetrized charge-weighted lengths defined by:

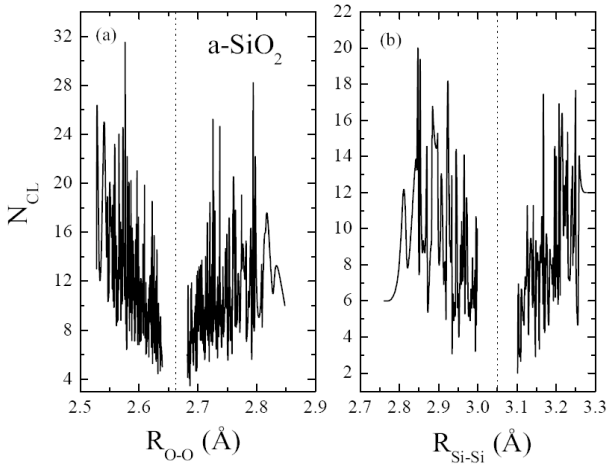

**Figure 7.** $N_{cl}$ as a function of second-neighbor distances (see text). Vertical lines are mean-distances.

$$R(E) = \frac{\sum_{l,m,n} \rho(l,n) q(l,E) q(m,E) q(n,E)}{\sum_{l,m,n} q(l,E) q(m,E) q(n,E)}, \quad (1)$$

where $\rho(l,n)$ is the distance between two second-neighbor sites $l$ and $n$ with shared site $m$ between them and $q$ is the charge computed from the electronic wave function (Kohn-Sham orbital) on a particular site for eigenenergy E. The mean deviation $\delta(E) = R(E) - R_0$, and $R_0$ is the average length for (either $R_{O-O}$ or $R_{Si-Si}$). In addition, the inverse participation ratio (IPR), a gauge of electronic localization, is shown in Fig. 8. An asymmetric contribution from the short $R_{O-O}$ and long $R_{Si-Si}$ lengths at the band tails is apparent. The localized valence tail states are clearly correlated to the short $R_{O-O}$ lengths and localized conduction tails associated with long Si-Si second neighbor distances.

The correlation between the denser volumes, characterized by the short $R_{O-O}$ lengths, and the valence tail states in a-SiO$_2$ system is reminiscent to the case of a-Si, and perhaps suggests a simple and broadly applicable understanding of the network origins of the Urbach edge, and the nature of the defect-induced strain field in disordered systems.

**6. Farther afield** There are indications that some of the effects we report here occur in (apparently) very different systems. Thus, in the important conjugated organic compound β-carotene ($C_{40}H_{56}$) a similar tendency is observed. In Fig. 9, we reproduce these results. The Highest Occupied Molecular Orbital (HOMO) state and Lowest Unoccupied Molecular Orbital (LUMO) state may be viewed as molecular analogues to the valence and conduction tails, respectively. The charge-weighted bond length $R(E)$ for the HOMO (LUMO) is 1.36 Å (1.44 Å) respectively.

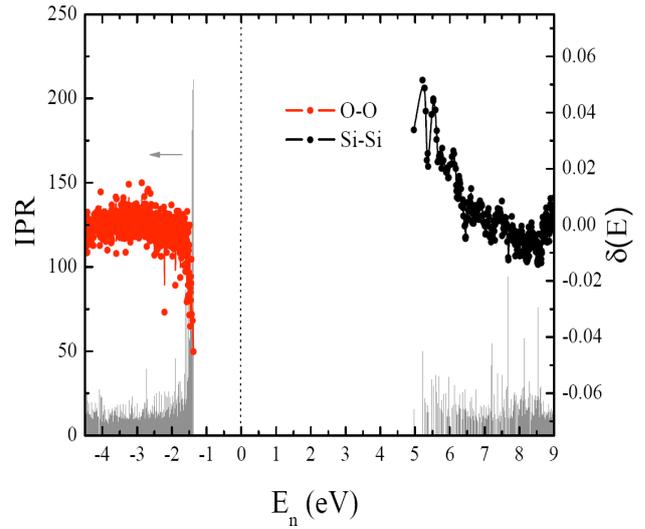

**Figure 8 (Color online)** Bond length decomposition of density of states. Dotted line indicates Fermi level. The valence (conduction) edge is dominated by short O-O bonds (long Si-Si bonds).

Both experiment and computation show that the conjugation length (*i.e.* the number of coplanar rings in a π-system, which in turn determines the overlap strength of the π orbitals) in carotene-β is 9.7, despite the fact that there are 11 double bonds in the molecule. The HOMO orbital is tied to the shortened double bonds. Preliminary re-sults on carotene-β suggest that frontier orbitals are local-ized at conjugation sites and may also contribute to lowering of the HOMO energy, compensating increase in energy due to the shortening of the double bonds (Fig. 9).

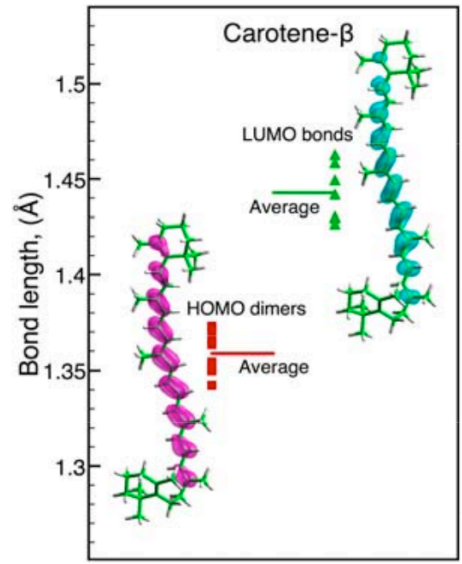

**Figure 9 (Color online)** HOMO (purple) and LUMO (cyan) states correspond to short and long bonds, with mean bond length 1.36Å (1.44Å) for HOMO (LUMO).

**7. Discussion** The systems we have discussed in this paper have only a few things in common. One is that all include disorder. A second feature is that all are relaxed to a minimum of suitable energy functionals (in this paper, always *ab initio*).

The consequence of a local densification around a short bond is quite plausible: the relaxation of the network around the nucleus is the strain field associated with the defect. Of course there is no similar statement for a long bond – a very long bond is a broken bond; the system does not become diffuse with long bonds, it reconstructs and forms voids.

In a-Si, the decay length for the spatial correlations from a nucleus is ~7Å. It is interesting that this distance is quite similar to the range of the density matrix in the material (for a review, see Ref. [8]). The latter is a measure of the spatial locality of quantum mechanics in a particular material, and also the range required of an accurate interatomic potential (specifically, the electronic contribution to the potential energy). Consistent with this observation, we note that the characteristic defect nucleus relaxation scale of a-$SiO_2$ is smaller, something like 5Å, consistent with the larger optical gap in silica, which is the prime determinant of the range of interatomic interactions.

The observation that short bond lengths are associated with valence tail states and long with conduction states is not a surprise. Within the usual tight-binding picture, the valence states are of bonding character and the conduction states of anti-bonding character, which leads naturally to the qualitative association of shorter bonds with valence states and longer with conduction. The following points substantially clarify this intuitive view: (1) In a purely covalent material (a-Si) the function *R(E)* (Eq. 1) is remarkably symmetric about $E_f$[4,5]; (2) the closer to $E_f$, the shorter (longer) the associated bonds on the valence (conduction) side of the Fermi level; (3) The short or long bonds are spatially correlated.

In connection with point (3), one might suppose *a priori* that the short or long bonds are randomly distributed in space. If this was so, the wave functions just above or below $E_f$ would be rather strange, with weight randomly dispersed over the network. But minimizing the energy in quantum mechanics is famously a "balancing act": the electronic energy is the sum of kinetic and potential energy terms. Thus, we conclude from these calculations that the kinetic energy term, which always favours delocalization, succeeds in "connecting up" the long or short bonds, albeit in a ~1-D fashion for the former and a higher dimension for the latter. This suggests a "chicken and egg" problem: Does the network organize itself (with structural blobs and filaments) to enable the electrons to reduce the Laplacian term (and delocalize electrons), or is the existence of structural blobs and filaments a consequence of minimizing the total energy (ions+electrons)? As we indicated above, it seems that some simple empirical potentials (even Keating springs) lead to structural blobs and filaments. To the extent that the empirical potentials themselves are constructed to reproduce experimental facts about the system (which in turn depend upon the electronic contribution to the total energy), perhaps the answer is not altogether clear.

A key feature of electron localization is that it creates a link between an electronic energy and a structural entity in a model. Thus, we observe valence tail states (that is, energy eigenstates from a specific narrow energy range) localized on blob-like regions. An analogous statement applies to filaments for conduction tail states. This has the interesting consequence that it causes Figs. 7 and 8 to be related. In fact the quantity $N_{cl}$ is very much like a participation ratio. This paper shows that there is a close connection between $R_{O-O}$, $R_{Si-Si}$ and the electronic energy *E* of Fig. 8. Thus, we show for example that dense regions in amorphous silica will particularly affect the valence tail, and lower density volumes will preferentially impact the conduction tail.

From an electronic point of view the existence of blobs and filaments associated with the quantum localized-to-extended transition has been demonstrated in realistic calculations for a-Si [9]. These papers sought to characterize the localized to extended transition in a real material (with the disorder being computed from realistic structural models). They clearly showed the blob+filament nature of states near the Fermi level. The weakness of these calculations is that the models are relatively small (so far up to $10^4$ atoms), whereas the 3D Anderson model has been diagonalized for systems exceeding $10^7$ atoms[10]. For such large systems, it becomes possible to compute the fractal dimension for the critical eigenstates (the result is not far from D=1.3, though there continuing dispute about the exact value). Qualitatively, our work shows that on the conduction side of the Fermi level states are quite filamentary (D=1+) and on the valence side something distinctly higher, say D=3-. The value of the fractal dimension at criticality is unknown for a real system in 3D.

Finally, we note that Phillips' model [11] for cuprate superconductivity emphasizes the importance of the filamentary states of the type we have detected in a-Si and a-$SiO_2$. Interestingly, atomic-resolution tunnelling asymmetry experiments on the cuprates have observed states with topology qualitatively like what we report here [12][13]. Given the appearance of blobs and filaments (both topological and electronic) in the diverse range of systems described here, and the known importance of disorder on the properties of the cuprates, we think this work lends some support to Phillips' view.

**Acknowledgements** We thank the US Army Research Office and US Army Research Lab for supporting this work, accomplished under cooperative agreement W911NF-0-2-0026 within the MURI program. We thank S. Chakraborty for support-ing calculations on a-SiO$_2$, and Dr. J. C. Phillips for helpful comments and encouragement. DAD thanks the US NSF and the Leverhulme Trust (UK) for sabbatical support. We thank Prof. Emilio Artacho for providing us with a carefully optimized basis set for SiO$_2$, and Prof. S. R. Elliott for many conversations and hospitality.